\documentclass[prb,showpacs,draft,twocolumn,superscriptaddress]{revtex4}

\usepackage[final]{graphicx}
\usepackage{bm}
\begin{document}

\title{Local exchange-correlation vector potential with memory in
  Time-Dependent Density Functional Theory: the
  generalized hydrodynamics approach}

\author{I.~V.~Tokatly}
\thanks{On leave from Moscow Institute of Electronic Technology,
         Zelenograd, 103498, Russia}

\email{ilya.tokatly@physik.uni-erlangen.de}

\affiliation{Lerhrstuhl f\"ur Theoretische Festk\"orperphysik,
  Universit\"at Erlangen-N\"urnberg, Staudtstrasse 7/B2, 91058
  Erlangen, Germany}

\author{O.~Pankratov}

\affiliation{Lerhrstuhl f\"ur Theoretische Festk\"orperphysik,
  Universit\"at Erlangen-N\"urnberg, Staudtstrasse 7/B2, 91058
  Erlangen, Germany} 
\date{\today}

\begin{abstract} 
Using Landau Fermi liquid theory we derive a nonlinear non-adiabatic
approximation for the exchange-correlation (xc) vector potential
defined by the xc stress tensor. The stress tensor is a local
nonlinear functional of two basic variables - the displacement vector
and the second-rank tensor which describes the evolution of momentum
in a local frame moving with Eulerian velocity. For irrotational
motion and equilibrium initial state the dependence on the tensor
variable reduces to that on the metrics generated by dynamical
deformation of the system.
\end{abstract}


\maketitle
Density-Functional Theory (DFT) offers an alternative to the common
wave function or Green's functions methods for studying the
properties of inhomogeneous many-body systems. Whereas the equilibrium
DFT has become a standard computational tool in solid state, atomic,
molecular and nuclear physics \cite{DFT}, the Time Dependent DFT
(TDDFT) \cite{TDDFT} is still intensively developing. An explosion of
interest in TDDFT is not surprising since the traditional many-body methods
encounter enormous computational difficulties if one attempts to consider
realistic non-equilibrium systems. The central problem of any DFT is
to find a good approximation for the xc potential. In the static DFT the
construction of approximations usually starts from the Local
Density Approximation (LDA), which, by itself, often gives very good
results. A similar approximation in TDDFT is still lacking. The
commonly used Adiabatic LDA (ALDA) can be justified only for
systems with the energy gap $\Delta$ if the inverse characteristic time
of the process $t_{pr}^{-1}=\omega \ll \Delta$. This condition is
strongly violated in most practical cases, which makes a construction
of the non-adiabatic functionals highly desirable.

In the mid-nineties it was realized that any consistent incorporation
of non-adiabaticity in the xc potential inevitably destroys its
spatial locality \cite{Vignale1}. An important step was made by
Vignale and Kohn (VK) \cite{VK} who showed that in the linear response
regime switching of the basic variable from density $n$ to current
${\bf j}$ allows to construct a consistent local non-adiabatic
approximation. The VK approximation is exact for $Lk_{F}\gg 1$
(which is a standard quasiclassical LDA condition, where $L$ is the
length scale of the density variation and $k_{F}$ is the local Fermi
momentum), and $\omega L/v_{F}\gg 1$. Vignale, Ullrich and Conti (VUC)
\cite{VUC} realized that velocity ${\bf v}={\bf j}/n$ is a more
natural variable than the current, as it allows to cast the VK result
in a transparent visco-elastic form. Yet available up to now {\it
nonlinear} non-adiabatic functionals \cite{Dobson1,VUC} have been
designed phenomenologically taking into account a number of
constraints due to the harmonic potential theorem \cite{HPT}, zero
force \cite{Vignale1} and torque condition and the known VK linear
form.  However, all these requirements, which must certainly be met,
do not uniquely determine the nonlinear xc functional.

The non-adiabatic LDA is in fact closely related to the generalized
hydrodynamics, which we derived in Ref.~\onlinecite{TP} to remove some
deficiencies of the adiabatic Bloch's theory \cite{Bloch}. In this
paper we extend the hydrodynamics formalism to the full-interacting case
and derive the nonlinear non-adiabatic local approximation for the xc
vector potential. The hydrodynamical approximation becomes exact under
the VK conditions $Lk_{F}\gg 1$, $\omega L/v_{F}\gg 1$, plus the
additional requirement that $\varepsilon_{F}/\omega \gg 1$
($\varepsilon_{F}$ is the local Fermi energy), which is a temporal
analogue of the inequality $Lk_{F}\gg 1$. In the linear regime our
local xc functional transforms to the VUC form.  It also shows an
initial state dependence, inevitable in any non-linear TDDFT \cite{Burke1}.

Let us first formulate TDDFT in the form which is convenient for further
discussion. Consider a non-equilibrium many-body system with
Hamiltonian $\hat{H} = \hat{T} + \hat{U}_{\text{ext}} + \hat{W}$,
where $\hat{T}$ is the kinetic energy operator, $\hat{U}_{\text{ext}}
= \int d{\bf x}U_{ext}({\bf x},t)\hat{n}({\bf x})$ describes the
interaction with the external potential, and $\hat{W}$ is the
interaction Hamiltonian related to the interparticle interaction
$V_{{\bf x}-{\bf x'}}$.

The averaged equations of motion for density $\hat{n}$ and
current $\hat{\bf j}$ operators can be represented in the form of the
hydrodynamical conservation laws
\begin{eqnarray}
 D_{t}n &+& n \partial_{\mu}v_{\mu} =0,
\label{2}\\
 mnD_{t}v_{\mu} &+& \partial_{\nu}P_{\mu\nu} + n\partial_{\mu}U = 0,
\label{3}
\end{eqnarray}
where $D_{t}=\partial_{t} + v_{\nu}\partial_{\nu}$ is the co-moving
derivative and $U=U_{\text{ext}}+U_{\text{H}}$ is a sum of
external and Hartree potentials. Since the Hartree term is
singled out in Eq.~(\ref{3}), the stress tensor $P_{\mu\nu} =
T_{\mu\nu} +W_{\mu\nu}$ contains only the kinetic $T_{\mu\nu}$ and the xc
$W_{\mu\nu}$ parts \cite{note1}
\begin{eqnarray} \label{4}
T_{\mu \nu} &=& \frac{1}{2m}\langle
(\hat{\pi}^{*}_{\mu}\psi^{+})(\hat{\pi}_{\nu}\psi) +
(\hat{\pi}^{*}_{\nu}\psi^{+})(\hat{\pi}_{\mu}\psi) -
\frac{\delta_{\mu \nu}}{2}\nabla^{2}n\rangle \quad\\
W_{\mu \nu} &=& -\frac{1}{2}\int d{\bf x'} x'_{\mu}
\partial'_{\nu}V_{{\bf x'}}\int_{0}^{1}
G_{{\bf x}+\lambda{\bf x'};{\bf x}-(1-\lambda){\bf x'}}d\lambda,
\label{5}
\end{eqnarray}
where $\hat{\pi}_{\mu} = -i\partial_{\mu} - mv_{\mu}$ is the relative
momentum, $G_{{\bf x};{\bf x'}} = \langle\psi^{+}({\bf x})\hat{n}({\bf
x'})\psi({\bf x})\rangle -n({\bf x})n({\bf x'})$ is the pair
correlation function and the angular brackets stand for exact
many-body state average. We use the summation convention over repeated
indeces throughout the paper.

According to the first part of the Runge-Gross (RG) proof of TDDFT
\cite{RG} (see also \cite{Raj}), the many-body wave function $\Psi(t)$
with initial condition $\Psi(0)=\Psi_{0}$, is a functional of the
current density ${\bf j}$ or, equivalently, of the velocity ${\bf
v}$. Hence the stress tensor is a functional of the velocity and of
the initial many-body state $P_{\mu\nu}[\Psi_{0},{\bf v}]$. Therefore
Eqs.~(\ref{2}) and (\ref{3}) are formally closed and RG theorem can be
viewed as a proof of existence of the exact quantum hydrodynamics with
the memory of the initial many-body correlations \cite{RG}.

The important part of DFT is the Kohn-Sham (KS) construction, which,
in a dynamical case, can be introduced as follows. Let us consider a
system of noninteracting KS particles moving in the self-consistent
scalar $U = U_{\text{ext}} +U_{H}$ and vector ${\bf A}^{\text{xc}}$
potentials. The local conservation laws for this system take the form
\begin{eqnarray}
 D_{t}n &+& n \partial_{\mu}v_{\mu} =0,
\label{6}\\
 mnD_{t}v_{\mu} &+& \partial_{\nu}P^{S}_{\mu\nu} 
 - nF^{\text{xc}}_{\mu} + n\partial_{\mu}U = 0,
\label{7}
\end{eqnarray}
where ${\bf F}^{\text{xc}} = -\partial_{t}{\bf A}^{\text{xc}} + {\bf
v}\times(\nabla\times{\bf A}^{\text{xc}})$ is the force due to the xc
vector potential, and KS stress tensor $P^{S}_{\mu\nu}$ is defined by
Eq.~(\ref{4}) with $\hat{\pi}_{\mu} = -i\partial_{\mu} - mv_{\mu}
-A^{\text{xc}}_{\mu}$.  Comparing Eqs.~(\ref{6}), (\ref{7}) with
Eqs.~(\ref{2}), (\ref{3}) we find that the density $n$ and the
velocity ${\bf v}$ of KS and interacting systems coincide if the
functional ${\bf A}^{\text{xc}}[n,{\bf v}]$ satisfies the equation
\begin{eqnarray}
F^{\text{xc}}_{\mu}=-\partial_{t} A^{\text{xc}}_{\mu} 
&+& ({\bf v}\times(\nabla\times{\bf A}^{\text{xc}}))_{\mu} =
 -\frac{1}{n}\partial_{\nu}\Delta P^{\text{xc}}_{\mu\nu},
\label{8}\\
\Delta P^{\text{xc}}_{\mu\nu} &=& P_{\mu\nu} - P^{S}_{\mu\nu}
\label{8a}
\end{eqnarray}
Equation (\ref{8}) is exactly of the form suggested by VUC \cite{VUC}
to satisfy Newton's third law \cite{note2}. Apparently ${\bf
A}^{\text{xc}}$ is a functional of the interacting initial state
$\Psi_{0}$ (via $P_{\mu\nu}$) and the KS initial state $\Psi^{S}_{0}$
(via $P^{S}_{\mu\nu}$) (for a recent discussion of the initial state
dependence see Ref.~\onlinecite{Burke1}).

Let us now turn to the local approximation for ${\bf
A}^{\text{xc}}$. Equation (\ref{8}) reduces the problem to the
construction of approximation for the xc stress tensor $\Delta
P^{\text{xc}}_{\mu\nu}$. As usual in LDA, we require that $Lk_{F}\gg
1$. In dynamics it is natural to assume the time-analogue of this
condition: $\varepsilon_{F}/\omega \gg 1$. At low temperatures the two
conditions bring us to the domain of applicability of Landau
Fermi-liquid theory \cite{Baym}. The latter is defined by the energy
functional $E[n_{\bf p}]$ (local in space and time) of the
quasiparticle distribution function $n_{\bf p}({\bf x},t)$. An
explicit example of this (in general nonlinear) functional is given by
the x-only approximation
\begin{equation}
E_{\text{x}}[n_{\bf p}] = \sum_{\bf p}\frac{{\bf p}^{2}}{2m}n_{\bf p} 
- \frac{1}{2}\sum_{\bf p,p'}V_{{\bf p}-{\bf p'}}n_{\bf p}n_{\bf p'}
\label{9}
\end{equation}
The function $n_{\bf p}$ satisfies the Boltzmann-type kinetic equation
\cite{Baym}. It is convenient to separate the convective motion with the
velocity ${\bf v}$ and to introduce the distribution function of relative
momentum $\tilde n_{\bf p} = n_{{\bf p} + m{\bf v}}$, which satisfies
the following equation \cite{TP}
\begin{eqnarray} \nonumber
D_{t}\tilde n_{\bf p}
 + \frac{\partial \varepsilon_{\bf p}}{\partial p_{\nu}}
   \frac{\partial \tilde n_{\bf p}}{\partial x_{\nu}}
 &-& \Big[
      mD_{t}v_{\nu} + p_{\mu}\frac{\partial v_{\nu}}{\partial x_{\mu}}\\
 &+& \frac{\partial \varepsilon_{\bf p}}{\partial x_{\nu}}
    + \frac{\partial U}{\partial x_{\nu}}
   \Big]
   \frac{\partial \tilde n_{\bf p}}{\partial p_{\nu}} 
  = I[\tilde n_{\bf p}]
\label{10}
\end{eqnarray}      
with initial condition $\tilde n_{\bf p}({\bf x},0)= \tilde N({\bf
p},{\bf x})$. Here $I[\tilde n_{\bf p}]$ is a collision integral, and
the quasiparticle energy is defined as $\varepsilon_{\bf p}=\delta
E/\delta \tilde n_{\bf p}$. Equation (\ref{10}) differs from the
common kinetic equation \cite{Baym} in two points. First, the time
derivative $\partial_{t}$ is replaced by the co-moving derivative
$D_{t}$ and, second, the force term contains additional inertial
contributions (the first two terms in the square brackets). The latter
arise due to the transformation to the local frame that moves with
Eulerian velocity ${\bf v}$. The first term $ mD_{t}v_{\nu}$ is the
linear acceleration force whereas the second term
$p_{\mu}\partial_{\mu} v_{\nu}$ is a sum of a force due to the
deformation rate
$\frac{1}{2}p_{\mu}(\partial_{\nu}v_{\mu}+\partial_{\mu}v_{\nu})$ and
the Coriolis force
$\frac{1}{2}p_{\mu}(\partial_{\mu}v_{\nu}-\partial_{\nu}v_{\mu})$.

Equation (\ref{10}) leads to the local conservation laws of
Eqs.~(\ref{2}), (\ref{3}) with the stress tensor expressed in
terms of the Landau functional \cite{Baym}
\begin{equation}
P_{\mu\nu} = \sum_{\bf p}p_{\mu} 
\frac{\partial \varepsilon_{\bf p}}{\partial p_{\nu}}\tilde n_{\bf p}
+ \delta_{\mu\nu} 
\Big[\sum_{\bf p}\varepsilon_{\bf p}\tilde n_{\bf p}
- E\Big].
\label{11}
\end{equation}
In the x-only approximation Eq.~(\ref{11}) reads
\begin{equation}
P^{\text{x}}_{\mu\nu} = \sum_{\bf p}
\frac{p_{\mu}p_{\nu}}{m}\tilde n_{\bf p}
- \sum_{\bf p,p'} \Big[p_{\mu} 
  \frac{\partial V_{{\bf p}-{\bf p'}}}{\partial p_{\nu}}
  + \frac{\delta_{\mu\nu}}{2}V_{{\bf p}-{\bf p'}} \Big]
\tilde n_{\bf p}\tilde n_{\bf p'}
\label{12}
\end{equation}
Substituting $\partial_{\nu}U$ from Eq.~(\ref{3}) into Eq.~(\ref{10}) we
eliminate the explicit dependence on the external potential from the
kinetic equation:
\begin{equation}
D_{t}\tilde n_{\bf p}
 + \frac{\partial \varepsilon_{\bf p}}{\partial p_{\nu}}
   \frac{\partial \tilde n_{\bf p}}{\partial x_{\nu}}
 - \left[
     p_{\mu}\frac{\partial v_{\nu}}{\partial x_{\mu}}
 + \frac{\partial \varepsilon_{\bf p}}{\partial x_{\nu}}
 - \frac{\partial P_{\mu\nu}}{n\partial x_{\mu}} 
   \right]
   \frac{\partial \tilde n_{\bf p}}{\partial p_{\nu}} = I
\label{13}
\end{equation}      
Equations (\ref{13}) and (\ref{11}) taken together with the definition of
the density $n = \sum_{\bf p}\tilde n_{\bf p}$ constitute a closed {\it
universal} (i.e. independent of external potential) problem. For a given
velocity ${\bf v}({\bf x},t)$ and initial conditions, this problem defines a
universal functional $P_{\mu\nu}[N({\bf p},{\bf x}),{\bf v}]$.

Calculation of the xc stress tensor $\Delta P^{\text{xc}}_{\mu\nu}$ of
Eq.~(\ref{8a}) requires a knowledge of the KS stress tensor
$P^{S}_{\mu\nu}$. The latter is obtained by the repetition of the
above consideration for the KS system. The resulting noninteracting
universal problem is
 \begin{eqnarray} \nonumber
D_{t}\tilde n^{S}_{\bf p}
 + \frac{p_{\nu}}{m}
   \frac{\partial \tilde n^{S}_{\bf p}}{\partial x_{\nu}}
 &-& \Big[
    p_{\mu}\frac{\partial v_{\nu}}{\partial x_{\mu}}
 - ({\bf p}\times (\nabla\times{\bf A}^{\text{xc}}))_{\nu}\\
    &-& \frac{\partial P^{S}_{\mu\nu}}{n\partial x_{\mu}}
   \Big]
   \frac{\partial \tilde n^{S}_{\bf p}}{\partial p_{\nu}} = 0,
\label{14}
\end{eqnarray}      
where $\tilde n^{S}_{\bf p}$ is a distribution function of KS
particles and the KS stress tensor is defined as follows
\begin{equation}
P^{S}_{\mu\nu} = \sum_{\bf p}
\frac{p_{\mu}p_{\nu}}{m}\tilde n^{S}_{\bf p}.
\label{15}
\end{equation}
Equations Eqs.~(\ref{13}), (\ref{11}), (\ref{14}), (\ref{15}), and
Eqs.~(\ref{8}), (\ref{8a}) allow to construct the dynamical counterpart of LDA,
provided the Landau functional $E[n_{\bf p}]$ is known.

Since the exchange self-energy is local in time, in x-only
case the collision integral in Eq.~(\ref{13}) vanishes, and
we can ignore the condition
$\varepsilon_{F}/\omega\gg 1$. Thus, irrespectively of the
frequency, ${\bf A}^{\text{x}}$ obtained with the Landau functional of
Eq.~(\ref{9}) is exact if $Lk_{F}\gg 1$. This is a direct dynamical
analogue of a static x-only LDA potential $v^{LDA}_{x}$, which
provides the long wave length limit of the time-dependent Optimized
Effective Potential (OEP) approximation.

In general, Eqs.~(\ref{13}), (\ref{11}) seem to be intractable. They
can be, however, solved in some important limiting cases. In a
collision-dominated regime $\nu_{\text{c}}/\max\{\omega,v_{F}/L\}\gg
1$ ($\nu_{\text{c}}$ is a collision frequency) the memory is rapidly
lost and the solution recovers the classical Navier-Stokes form of
$P_{\mu\nu}$ \cite{LandauVI}. On the contrary, the memory effects are
pronounced in the opposite limit of a fast collisionless motion. In
this regime the solution to Eqs.~(\ref{13}), (\ref{11}) can be found
analytically if $L\omega/v_{F}\gg 1$, which is the VK condition
\cite{VK} and, at the same time, the condition of applicability of the
generalized hydrodynamics \cite{TP}. In the rest of the paper we 
analyze this case in detail and derive the local non-adiabatic
approximation for $\Delta P^{\text{xc}}_{\mu\nu}$.
 
Let us estimate different terms in Eq.~(\ref{13}). The time-derivative
term and the first term in brackets in Eq.~(\ref{13}) are of the order
of $\omega$ (the continuity equation requires $v/L \sim\omega$),
whereas all other terms give contributions $\sim v_{F}/L$. Hence to
the leading order in $v_{F}/\omega L \ll 1$ only inertial forces are
relevant, and, therefore, the function $\tilde n_{\bf p}$ satisfies
the following equation
\begin{equation}
\left[ \frac{\partial}{\partial t}
+ v_{\nu}\frac{\partial}{\partial x_{\nu}}
    - p_{\mu}\frac{\partial v_{\nu}}{\partial x_{\mu}}
   \frac{\partial}{\partial p_{\nu}}\right]\tilde n_{\bf p}({\bf x},t) = 0
\label{16}
\end{equation}
with initial condition $\tilde n_{\bf p}({\bf x},0)=\tilde N({\bf
 p},{\bf x})$. In Eq.~(\ref{16}) we used an explicit
 expression for the co-moving derivative $D_{t}$. With the same
 accuracy, the KS distribution function $\tilde n^{S}_{\bf p}({\bf
 x},t)$, which defines the KS stress tensor $P^{S}_{\mu\nu}$ of
 Eq.~(\ref{15}), also satisfies Eq.~(\ref{16}), but with another
 initial condition $\tilde n^{S}_{\bf p}({\bf x},0)=\tilde N^{S}({\bf
 p},{\bf x})$.

To solve Eq.~(\ref{16}) we introduce a nonlinear transformation of
variables ${\bf x},{\bf p} \to{\bm \xi},{\bf k}$: 
\begin{eqnarray} \label{18}
{\bf x} &=& {\bf x}({\bm \xi},t)={\bm \xi} + {\bf u}({\bm \xi},t),\\
{\bf p} &=& {\bf p}({\bm \xi},{\bf k},t),
 \label{19}
\end{eqnarray}
where the functions ${\bf x}({\bm \xi},t)$ and ${\bf p}({\bm \xi},{\bf
  k},t)$ are solutions to the following initial value problems
\begin{eqnarray} \label{20}
\frac{\partial {\bf x}(t)}{\partial t} &=& {\bf v}({\bf x}(t),t), 
        \quad {\bf x}(0)= {\bm \xi}\\
\frac{\partial p_{\mu}(t)}{\partial t} &=&
 - \frac{\partial v_{\mu}({\bf x}(t),t)}{\partial x_{\nu}} p_{\nu}(t),
        \quad  p_{\mu}(0) = k_{\mu}.
 \label{21}
\end{eqnarray}
In terms of the new variables Eq.~(\ref{16}) transforms to the equation
$\partial_{t}\tilde n({\bf k},{\bm \xi},t)=0$, which is trivially
solvable. Hence the solution of the original problem is
\begin{equation}
\tilde n_{\bf p}({\bf x},t) = 
\tilde N({\bf k}({\bf x},{\bf p},t),{\bm \xi}({\bf x},t)),
\label{22}
\end{equation}
where ${\bm \xi}({\bf x},t)$ and ${\bf k}({\bf x},{\bf p},t)$ are
obtained by inversion of Eqs.~(\ref{18}) and (\ref{19}). Variable
${\bm \xi}$, Eq.~(\ref{18}), is a Lagrangian coordinate
\cite{acoustics} which has a meaning of the initial position of the
fluid element presently at ${\bf x}$. Vector ${\bf u} = {\bm x} - {\bm
\xi}$ is thus a displacement of the fluid element. Similarly, ${\bf
k}$ is the initial momentum of a quasiparticle initially at ${\bm
\xi}$ that moves under inertial forces and, at the time moment $t$,
acquires a momentum ${\bf p}$.

In the Eulerian (spatial) description \cite{acoustics} the above
solution for the quasiparticle $\tilde n_{\bf p}({\bf x},t)$ and KS
$\tilde n^{S}_{\bf p}({\bf x},t)$ distribution functions can be
written as
\begin{eqnarray}\label{23}
\tilde n_{\bf p}({\bf x},t) &=& 
\tilde N(p_{\nu}\eta^{-1}_{\mu\nu}({\bf x},t),{\bf x}-{\bf u}({\bf x},t)),\\
\tilde n^{S}_{\bf p}({\bf x},t) &=& 
\tilde N^{S}(p_{\nu}\eta^{-1}_{\mu\nu}({\bf x},t),{\bf x}-{\bf u}({\bf x},t)),
\label{24}
\end{eqnarray}
where the displacement vector ${\bf u}({\bf x},t)$ and the second-rank
tensor $\eta_{\mu\nu}({\bf x},t)$ are defined as follows
\begin{eqnarray}\label{25}
D_{t}{\bf u}({\bf x},t) &=& {\bf v}({\bf x},t),
 \quad  {\bf u}({\bf x},0) = 0,   \\
D_{t}\eta_{\mu\nu}({\bf x},t) &=& - 
\frac{\partial v_{\mu}}{\partial x_{\alpha}}\eta_{\alpha\nu}({\bf x},t), 
 \text{  } \eta_{\mu\nu}({\bf x},0)=\delta_{\mu\nu}
\label{26}
\end{eqnarray}
Tensor $\eta_{\mu\nu}$ relates the momenta ${\bf p}$ and ${\bf k}$
and, therefore, describes the rotation and the stretching of a
quasiparticle momentum, which are caused by the inertial forces in the
frame moving with Eulerian velocity. Using Eq.~(\ref{26}) one can
prove that this tensor has an important property - the determinant of
the matrix $\eta_{\mu\nu}$ equals to the Jacobian $J$ of the
coordinate transformation of Eq.~(\ref{18}). Hence $\det\eta_{\mu\nu}$
is directly related to the density $n({\bf x},t)$:
\begin{equation}
\det\eta_{\mu\nu} = \det \frac{\partial\xi_{\mu}}{\partial x_{\nu}}
= J({\bf x},t) = \frac{n({\bf x},t)}{n_{0}({\bm \xi}({\bf x},t))},
\label{27}
\end{equation}  
where $n_{0}({\bf x})$ is the initial density distribution. In a
particular case of irrotational motion the Coriolis force vanishes
($\partial_{\mu}v_{\nu}=\partial_{\nu}v_{\mu}$), and the solution to
Eq.~(\ref{26}) is locally expressed in terms of deformation gradients
\begin{equation}
\eta_{\mu\nu} = \frac{\partial\xi_{\nu}}{\partial x_{\mu}} =
\delta_{\mu\nu} - \frac{\partial u_{\nu}}{\partial x_{\mu}}.
\label{28}
\end{equation}
Substituting the distribution functions Eqs.~(\ref{23}) and (\ref{24})
into the definitions of the stress tensors $P_{\mu\nu}$,
Eq.~(\ref{11}), and $P^{S}_{\mu\nu}$, Eq.~(\ref{15}), and calculating
their difference, Eq.~(\ref{8a}), we obtain the final expression for
$\Delta P^{\text{xc}}_{\mu\nu}$. The xc stress tensor $\Delta
P^{\text{xc}}_{\mu\nu}$ is a local functional (function) of two basic
variables ${\bm \xi}({\bf x},t)= {\bf x} -{\bf u}({\bf x},t)$ and
$\eta_{\mu\nu}({\bf x},t)$
\begin{equation}
\Delta P^{\text{xc}}_{\mu\nu}({\bf x},t))= \Delta
P^{\text{xc}}_{\mu\nu}(\eta_{\mu\nu}({\bf x},t),{\bm \xi}({\bf x},t)).
\label{29}
\end{equation}
It is also a functional of the initial distribution functions $\tilde
N({\bf p},{\bf x})$ and $\tilde N^{S}({\bf p},{\bf x})$ as it must be
for any consistent approximation in TDDFT. If the evolution starts
from the ground state we have $\tilde N=\tilde N^{S}= f^{F}_{\bf
p}({\bf x})= \theta(k_{F}^{2}({\bf x})-{\bf p}^{2})$, and the initial
state dependence reduces to the dependence on the initial density
$n_{0}({\bf x})$ (here $k_{F}({\bf x})$ is the local Fermi momentum at
$t=0$). In addition, the isotropy in ${\bf p}$-space of the
equilibrium initial distribution function requires that $\Delta
P^{\text{xc}}_{\mu\nu}$ should not depend directly on $\eta_{\mu\nu}$,
but only via the symmetric tensor
$g_{\mu\nu}=\eta_{\mu\alpha}\eta_{\nu\alpha}$:
\begin{equation}
\Delta P^{\text{xc}}_{\mu\nu}({\bf x},t))=
\Delta P^{\text{xc}}_{\mu\nu}
(g_{\alpha\gamma}({\bf x},t),n_{0}({\bm \xi}({\bf x},t))).
\label{30}
\end{equation}

Interestingly, in a special case of irrotational motion, using
Eq.~(\ref{28}) we obtain
$
g_{\mu\nu}({\bf x},t)=\frac{\partial\xi_{\alpha}}{\partial x_{\mu}}
\frac{\partial\xi_{\alpha}}{\partial x_{\nu}},
$ 
which is exactly the
metrics generated by the coordinate transformation of Eq.~(\ref{18}): 
$(d{\bm \xi})^{2} = g_{\mu\nu}dx_{\mu}dx_{\nu}$.

In the linear response regime Eq.~(\ref{23}) leads to the following
expression for the deviation of the distribution function 
$\tilde n_{\bf p}$ from its initial value $\tilde N({\bf p},{\bf x})$
\begin{equation}
\delta \tilde n_{\bf p} = 
- u_{\mu}\frac{\partial}{\partial x_{\mu}}\tilde N({\bf p},{\bf x}) 
- p_{\nu }\frac{\partial u_{\mu}}{\partial x_{\nu}}
\frac{\partial}{\partial p_{\mu}}\tilde N({\bf p},{\bf x}),
\label{31}
\end{equation}
Substituting Eq.~(\ref{31}) into Eq.~(\ref{11}) and considering the
equilibrium initial distribution $\tilde N({\bf p},{\bf x})=f^{F}_{\bf
p}({\bf x})$, we recover the correct elastic form of the stress tensor
\cite{VUC}. We note, that viscosity vanishes within the limits of
applicability of Landau theory.

The expression for the xc stress tensor Eq.~(\ref{30}) obviously
satisfies all general theorems and limiting requirements, as initial
microscopic equations do. It is noteworthy that the dependence on a
delayed (in fact, Lagrangian) coordinate in Eq.~(\ref{30}) radically
differs from the form suggested in Refs.~\onlinecite{Dobson1,VUC} on
phenomenological grounds.

To illustrate Eq.~(\ref{30}), which is one of the main results
of the paper, we present an explicit expression for the x-only stress
tensor
\begin{equation}
\Delta P^{\text{x}}_{\mu\nu} = 
- \frac{J}{2}\sum_{\bf p} V_{\bf p} p_{\mu}
\frac{\partial}{\partial p_{\nu}} G^{\text{x}}
(\eta_{\alpha\gamma}p_{\gamma},n_{0}({\bm \xi})),
\label{32}
\end{equation}
where 
$G^{\text{x}}({\bf p},n) = -\sum_{\bf k}f^{F}_{\bf k}f^{F}_{\bf k+p}$
is the exchange pair correlation function in a homogeneous system with
the density $n$. For a 3D system with Coulomb interaction from Eq.~(\ref{32})
follows
$$
\Delta P^{\text{x}}_{\mu\nu} = -n\frac{e^{2}k_{F}({\bm \xi})}{\pi^{2}}
\int \frac{d^{3}{\bf q}}{q^{2}}
\Big[\frac{\delta_{\mu\nu}}{2} - \frac{q_{\mu}q_{\nu}}{q^{2}}\Big]
F\left(\sqrt{g_{\alpha\gamma}^{-1}q_{\alpha}q_{\gamma}}\right)
$$
where $F(y)=[1 - 3y/2 + y^{3}/2]\theta(1-y)$. It important to note
that this equation (and more general Eq.~(\ref{32})) is, in fact, the
explicit small-gradient limit of the dynamic OEP functional. 

A construction of the explicit form of $\Delta
P^{\text{xc}}_{\mu\nu}$ beyond the x-only approximation requires a
knowledge of the Landau functional or, more precisely, the functional
$P_{\mu\nu}[n_{\bf p}]$ of Eq.~(\ref{11}) for a homogeneous
system. This is obviously in a spirit of LDA, which treats an
inhomogeneous system as a locally homogeneous one. In principle, it
should be possible to calculate $P_{\mu\nu}[n_{\bf p}]$ using the
non-equilibrium Keldysh technique \cite{Keldysh}. As is evident
from Eq.~(\ref{11}), the tensor $P_{\mu\nu}[n_{\bf p}]$ is uniquely
defined by the quasiparticle energies $\varepsilon_{\bf p}$ and the
total energy $E$. The both quantities are the functionals of the full
Green's function $G$. Using the general reconstruction formulas
\cite{Kremp} one can relate $G$ to the quasiparticle distribution
$n_{\bf p}$ and thus restore the required functional. If the system
evolution starts from the ground state, the derived dynamical local
approximation actually requires the functional $P_{\mu\nu}[n_{\bf p}]$
only on a set of $t$- and ${\bf x}$-independent functions $n_{\bf p}$
which describe a Fermi-type occupation of a volume of the ${\bf
p}$-space that is enclosed in a second-order surface, as given by
Eq.~(\ref{23}) for $\tilde N({\bf p},{\bf x})=\theta(k_{F}^{2}({\bf
x})-{\bf p}^{2})$. This restriction should essentially simplify a
calculation of $P_{\mu\nu}[n_{\bf p}]$, which still remains a 
demanding, though feasible, task.

In conclusion, we derived the nonlinear non-adiabatic counterpart of
LDA. We showed that the xc stress tensor (and consequently the xc vector
potential) is a local functional of one vector and one tensor
variables, which describe dynamical deformation of coordinate and
momentum space respectively.

I.T. is grateful to the
Alexander von Humboldt Foundation for support.

\end{document}